\documentclass[aps,prb,twocolumn,showpacs,groupedaddress]{revtex4}

\usepackage{graphicx}
\usepackage{latexsym}
\usepackage{amssymb}

\begin{document}


\title{Stretched exponential relaxation in three dimensional short-range spin glass Cu$_{0.5}$Co$_{0.5}$Cl$_{2}$-FeCl$_{3}$ graphite bi-intercalation compound}

\author{Itsuko S. Suzuki }
\email[]{itsuko@binghamton.edu}
\affiliation{ Department of Physics, State University of New York
at Binghamton, Binghamton, New York 13902-6000}

\author{Masatsugu Suzuki }
\email[]{suzuki@binghamton.edu}
\affiliation{ Department of Physics, State University of New York
at Binghamton, Binghamton, New York 13902-6000}


\date{\today}

\begin{abstract}
Cu$_{0.5}$Co$_{0.5}$Cl$_{2}$-FeCl$_{3}$ graphite bi-intercalation compound is a three-dimensional short-range spin glass with a spin freezing temperature $T_{SG}$ ($= 3.92 \pm 0.11$ K). The time evolution of the zero-field cooled magnetization $M_{ZFC}(t)$ has been measured under various combinations of wait time ($t_{w}$), temperature ($T$), temperature-shift ($\Delta T$), and magnetic field ($H$). The relaxation rate $S_{ZFC}(t)$ [$=(1/H)$d$M_{ZFC}(t)$/d$\ln t$] shows a peak at a peak time $t_{cr}$. The shape of $S_{ZFC}(t)$ in the vicinity of $t_{cr}$ is well described by stretched exponential relaxation (SER). The SER exponent $b$ and the SER relaxation time $\tau_{SER}$ are determined as a function of $t_{w}$, $T$, $H$, and $\Delta T$. The value of $b$ at $T=T_{SG}$ is nearly equal to 0.3. There is a correlation between $\tau_{SER}$ and $1/b$, irrespective of the values of $t_{w}$, $T$, $H$, and $\Delta T$. These features can be well explained in terms of a simple relaxation model for glassy dynamics.
\end{abstract}


\maketitle

\section{\label{intro}Introduction}
Recently the aging phenomena have been the subject of many experimental studies on slow dynamics in a variety of spin glass (SG) systems.\cite{ref01,ref02,ref03,ref04} Typically it can be observed in the evolution of a zero-field cooled (ZFC) magnetization $M_{ZFC}(t)$ with time $t$ after the ZFC aging protocol for a wait time $t_{w}$. The aging behavior can be understood based on a phenomelogical domain model.\cite{ref03,ref05,ref06} In this picture, the aging involves the growth of the domain (denoted by $R$) during the ZFC protocol for a wait time $t_{w}$. The domain grows with time. The size of the domain $R$ becomes equal to $R(t_{w})$ after the wait time $t_{w}$. Through this process, only the relaxation time $\tau$, which is nearly equal to $t_{w}$, can be selected. At $t=0$ just after the ZFC protocol, a magnetic field is turned on. Then the ZFC magnetization is measured as a function of the observation time $t$. The size of the domain ($R$) remains constant $R(t_{w})$ for $0<t<t_{w}$. In contrast, the probing length scale ($L$) of the domain grows with the time $t$, starting from $t=0$ in a similar way such that the domain (size $R$) grows for the wait time $t_{w}$ during the ZFC protocol. The equilibrium dynamics is probed since $L<R(t_{w})$ for $0<t<t_{w}$, while the non-equilibrium dynamics is probed for $t>t_{w}$. 

An usual way to describe the slow relaxation of the ZFC magnetization is to postulate a statistical distribution of the relaxation times and to assume additive contributions. According to Lundgren et al,\cite{ref01,ref02} the ZFC magnetization $M_{ZFC}(t_{w},t)$ is described by a sum of exponential decay $\exp(-t/\tau)$ with the relaxation time $\tau$ multiplied by the density of relaxation times $g(t_{w},\tau)$, 
\begin{eqnarray} 
\frac{1}{H}[M_{ZFC}(t_{w},t)-M_{0}]=q(t_{w},t)  \nonumber \\
=-\int_{\tau_{0}}^{\infty}g(t_{w},\tau)\exp(-\frac{t}{\tau})d\tau,
\label{EQN01} 
\end{eqnarray} 
where $H$ is the magnitude of an external magnetic field, $M_{0}$ is the ZFC magnetization at $t$ = 0, and $\tau_{0}$ is a microscopic relaxation time ($\tau_{0}\simeq 10^{-12}$ sec). The relaxation rate $S_{ZFC}(t_{w}, t)$ can be defined as
\begin{eqnarray} 
S_{ZFC}(t_{w},t)=\frac{1}{H}\frac{dM_{ZFC}(t_{w},t)}{d\ln t}
=\frac{dq(t_{w},t)}{d\ln t}  \nonumber \\
=\int_{\tau_{0}}^{\infty}g(t_{w},\tau)\frac{t}{\tau}\exp(-\frac{t}{\tau})d\tau .
\label{EQN02}
\end{eqnarray}
Here it is noted that a part of the integrand expressed by $f(x)=(1/x)\exp(-1/x)$ has a maximum at $x = 1.0$, where $x=\tau /t$. Using an assumption that $f(\tau /t)$ is approximated by a Dirac-delta function [$=\delta(t-\tau)$],\cite{ref01,ref02,ref04} we get
\begin{equation} 
S_{ZFC}(t_{w},t)\approx\int_{\tau_{0}}^{\infty}g(t_{w},\tau)\delta(t-\tau)d\tau
=g(t_{w},t) .
\label{EQN03}
\end{equation}

Experimentally it is well known that $S_{ZFC}(t, t_{w})$ has a relatively flat peak centered around $t=t_{w}$. This implies that the density of the relaxation time $g(t_{w},\tau)$ also exhibits a broad peak around $\tau=t_{w}$ because of $g(t_{w}, \tau)=S_{ZFC}(t_{w},\tau)$, reflecting the glassy state. Here the label $t_{w}$ is used for the notation of $M_{ZFC}(t_{w}, t)$ and $g(t_{w}, t)$, in order to emphasize that each relaxation rate $S_{ZFC}(t_{w}, t)$ [$\approx g(t_{w}, t)$] after the ZFC protocol for the wait time $t_{w}$ represents the dominant feature of the aging dynamics for a specific domain which grows for the wait time $t_{w}$ during the ZFC protocol. A set of data on $S_{ZFC}(t_{w}, t)$ as a function of $t$, for various $t_{w}$ (=$10^{2}-10^{5}$ sec) provides an information on the aging behavior of domains whose size depends on the wait time. For short $t_{w}$, one can get the aging behavior for small domains, while for long $t_{w}$, one can get the aging behavior for large domains. The relaxation mechanism for the small size domains is considered to be rather different from that for the large size domain. There is a crossover between the thermal-equilibrium dynamics inside the domains and the non-equilibrium dynamics in domain walls.

Cu$_{0.5}$Co$_{0.5}$Cl$_{2}$-FeCl$_{3}$ graphite bi-intercalation compound (GBIC) magnetically behaves like a 3D short-ranged SG. This compound undergoes a SG transition at $T_{SG}= 3.92 \pm 0.11$ K in the absence of $H$. In our previous papers,\cite{ref07,ref08,ref09} we have undertaken an extensive study on the aging behavior of the SG phase of our system from the time dependence of $M_{ZFC}(t)$ under various kinds of conditions where $t_{w}$, $T$, $H$ and $\Delta T$ (the $T$-shift) are changed as parameters (see Sec.~\ref{result} for the detail of experimental procedure). The relaxation rate $S_{ZFC}(t)$ shows a peak at a peak time $t_{cr}$. The value of $t_{cr}$ depends on the parameters $t_{w}$, $T$, $H$, and $\Delta T$. 

In the present paper, we show that the $t$ dependence of $S_{ZFC}(t)$ is well described by the SER form in the vicinity of $t\approx t_{cr}$, irrespective of $t_{w}$, $T$, $H$, and $\Delta T$. The peak time $t_{cr}$ is equal to the SER relaxation time $\tau_{SER}$. The least squares fit of the data in the vicinity of $t=t_{cr}$ to the SER form yields the SER relaxation time $\tau_{SER}$ and the SER exponent $b$, and the SER maximum $S_{max}^{SER}$. Our results are summarized by the following two features. (i) the SER exponent $b$ increases with increasing $T$. (ii) There is a strong correlation between $\tau_{SER}$ and $1/b$. These features are seen in many systems other than SG's, and is considered a signature of the glassy relaxation. We show that these features can be well explained in terms of a simple model of glassy relaxation.\cite{ref10,ref11}

\section{\label{back}Relaxation rate $S_{ZFC}(t_{w},t)$}
\subsection{\label{backA}General form}

\begin{figure}
\includegraphics[width=7.5cm]{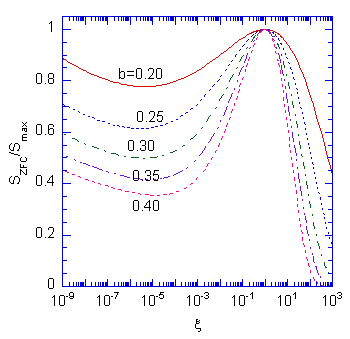}
\caption{\label{fig01}(Color online)Plot of $S_{ZFC}(t)/S_{max}$ vs $\xi = t/t_{cr}$ which is derived from the combination of Eqs.(\ref{EQN05}), (\ref{EQN06}), (\ref{EQN08}), and (\ref{EQN09}). $a = 0.035$. $b$ is changed as a parameter; $b$ = 0.20, 0.25, 0.30, 0.35, and 0.40. $t_{cr}$ is the peak time of $S_{ZFC}(t)$. $b>4a=0.14$.}
\end{figure}

Theoretically and experimentally it has been accepted that the the time variation of $M_{ZFC}(t_{w},t)$ may be described by\cite{ref12}
\begin{equation} 
\frac{1}{H}[M_{ZFC}(t_{w},t)-M_{0}]=-At^{-a}\exp[-(t/\tau)^{b}],
\label{EQN04}
\end{equation}
around $t\simeq t_{w}$, where $a$ ($a>0$) is called the pre-factor exponent and $b$ ($0<b<1$) is called the SER exponent, $A$ is a constant, and $t_{w}$ is the wait time. Then the relaxation rate $S_{ZFC}(t_{w},t)$ can be derived as
\begin{eqnarray} 
S_{ZFC}(t_{w},t)=\frac{1}{H}\frac{dM_{ZFC}(t_{w},t)}{d\ln t} \nonumber \\
=At^{-a}\exp[-(t/\tau)^{b}][a+b(t/\tau)^{b}] .
\label{EQN05}
\end{eqnarray}
$S_{ZFC}(t_{w},t)$ has a local maximum [$S_{max}$] at $t=t_{cr}$. The peak time $t_{cr}$ is given by
\begin{equation} 
\frac{t_{cr}}{\tau}=(\frac{1}{2}-\frac{a}{b}+\frac{\sqrt{b^{3}(b-4a)}}{2b^{2}})^{1/b},
\label{EQN06}
\end{equation}
under the condition that $b>4a$. When $a=0$, we have
\begin{equation} 
\frac{t_{cr}}{\tau}=1 .
\label{EQN07}
\end{equation}
The maximum $S_{max}$ is also given by
\begin{eqnarray} 
S_{max}=\frac{A}{b}2^{-1+a/b}[b^2+\sqrt{b^{3}(b-4a)}] \nonumber \\
\times [\frac{-2ab+b^{2}+\sqrt{b^{3}(b-4a)}}{b^{2}}]^{-a/b} \nonumber \\
\times \exp[\frac{-2ab+b^{2}+\sqrt{b^{3}(b-4a)}}{b^{2}}]\tau^{-a} .
\label{EQN08}
\end{eqnarray}
Figure \ref{fig01} shows the normalized relaxation rate defined by $S_{ZFC}/S_{max}$ as 
a function of the normalized time $\xi$ defined by
\begin{equation} 
\xi=\frac{t}{t_{cr}}=\frac{t/\tau}{t_{cr}/\tau} ,
\label{EQN09}
\end{equation}
where $a$ = 0.035 (which is appropriate for our system, see Sec.~\ref{result0}) and $b$ is changed as a parameter. The normalized relaxation rate $S_{ZFC}/S_{max}$ has a peak at $\xi=1$. The width of the peak in $S_{ZFC}/S_{max}$ vs $\xi$ becomes narrow when $b$ becomes increases.

\subsection{\label{backB}SER form for the least-squares fitting}
In the case of intermediate $t_{w}$, one can find the time dependence of the relaxation rate for the relaxation of the sufficiently large domain size. Since the power form $t^{-a}$ is a slowly varying function of $t$ because of small value $a$ (= 0.035) in our system, $M_{ZFC}(t_{w},t)$ is approximated by the pure SER form, 
\begin{equation}
M_{ZFC}(t_{w},t)=M_{0}-A\exp [-(\frac{t}{\tau_{SER}})^{b}] ,
\label{EQN10}
\end{equation}
in the narrow time regime near $t=t_{w}$, where the prefactor $a$ is equal to 0, $b$ is the SER exponent and $\tau_{SER}$ is the SER relaxation time and is nearly close to $t_{w}$. Then the relaxation rate $S_{ZFC}(t_{w},t)$ is expressed by 
\begin{equation}
S_{ZFC}(t_{w},t)=eS_{\max }^{SER}(\frac{t}{\tau_{SER}})^{b}\exp[-(\frac{t}{\tau_{SER}})^{b}],
\label{EQN11}
\end{equation}
where $e$ is the base of natural logarithmic. This function is almost symmetric with respect to the axis $t/\tau_{SER}$ = 1 in the logarithmic scale of the $t/t_{SER}$ axis. Since the density of the relaxation time $g(t_{w},\tau)$ is nearly equal to $S_{ZFC}(t_{w},\tau)$, $g(t_{w},\tau)$ has a broad peak in the very vicinity of $\tau=t_{w}$. In other words, a set of data on $S_{ZFC}(t_{w},t)$ as a function of $t$, for various $t_{w}$, provides an information on the aging behavior of only domains whose size depends on the wait time. Then the exponent $b$ should be determined from the least-squares fit of the data of $S_{ZFC}(t_{w},t)$ vs $t$ in the very vicinity of $t_{w}$. To this end, in the present analysis, typically we use the data for the limited time regime $0.1<t/t_{w}<10$. Note that technically it very difficult to determine both $a$ and $b$ from the least-squares fit of the data of $S_{ZFC}(t_{w},t)$ vs $t$ in the very vicinity of $t_{w}$. One of the reason is that $b$ is strongly dependent on the slight change of the exponent $a$. 

\section{\label{exp}EXPERIMENTAL PROCEDURE}
The detail of sample characterization and sample preparation of Cu$_{0.5}$Co$_{0.5}$Cl$_{2}$-FeCl$_{3}$ GBIC was provided in our previous papers.\cite{ref07,ref08,ref09} The DC magnetization was measured using a SQUID magnetometer (Quantum Design, MPMS XL-5) with an ultra low field capability option. The remnant magnetic field was reduced to zero field (exactly less than 3 mOe) at 298 K. The time ($t$) dependence of the zero-field cooled (ZFC) magnetization ($M_{ZFC}$) was measured. The following ZFC aging protocol was carried out before the measurement. First the sample was annealed at 50 K for $1.2\times 10^{2}$ sec in the absence of $H$. Then the system was quenched from 50 K to $T$ ($<T_{SG}$). It was aged at $T$ for a wait time $t_{w}$ (typically $t_{w}=2.0\times 10^{3}$ - $3.0\times 10^{4}$ sec). After the wait time, an external magnetic field $H$ is applied along any direction perpendicular to the $c$-axis at $t$ = 0. The measurements of $M_{ZFC}$ vs $t$ were made under the various conditions. The detail of the experimental procedure will be described in Sec.~\ref{result}.

\section{\label{result}RESULT}
\subsection{\label{result0}Determination of the pre-factor exponent $a$}
In the early stage of the time evolution ($t\simeq 0$, $t_{w}\simeq 0$, but still $t\ll t_{w}$), the ZFC magnetization can be well described by a power-law form:
\begin{equation} 
\frac{1}{H}[M_{ZFC}(t_{w},t)-M_{0}]=-At^{-a} .
\label{EQN12}
\end{equation}
The corresponding relaxation rate is obtained as
\begin{equation} 
S_{ZFC}(t_{w},t)\approx t^{-a} .
\label{EQN13}
\end{equation}
The relaxation rate monotonically decreases with increasing $t$ in the early stage. In our measurement using SQUID magnetometer, it is very difficult to measure the time dependence of $M_{ZFC}(t_{w}, t)$ for very short $t_{w}$. No reliable data can be taken for small $t_{w}$, since it takes relatively long time until for the temperature to become stable before the ZFC process starts. There is some uncertainty for the definition of $t_{w}$. In the early stage ($t$ = 0), in turn, we use the frequency dependence of the absorption of the AC magnetic susceptibility to determine the value of $a$. The absorption $\chi^{\prime\prime}(\omega)$ of the AC magnetic susceptibility is related to the relaxation rate through a so-called $\pi/2$ law,\cite{ref13}
\begin{equation}
\chi^{\prime\prime}(\omega)=-\frac{\pi}{2}S_{ZFC}(t_{w}=1/\omega,t=0)\approx \omega^{a} .
\label{EQN14}
\end{equation}
In our previous paper\cite{ref07} we experimentally determine the value of $a$ as
\begin{equation}
a=0.035\pm 0.001 .
\label{EQN15}
\end{equation}
In fact, it is experimentally confirmed that this exponent $a$ is related to the critical exponents $b$, and $z\nu$ by 
\begin{equation}
a=\frac{\beta}{z\nu} ,
\label{EQN16}
\end{equation}
where $\beta$ is the critical exponent of the order parameter ($\beta=0.36\pm 0.03$), $z$ is the dynamic critical exponent, and $\nu$ is the critical exponent of the inverse correlation length in the present system; $z\nu = 10.3\pm 0.7$. 

\subsection{\label{resultA}Aging behavior at various $T$}

\begin{figure}
\includegraphics[width=7.5cm]{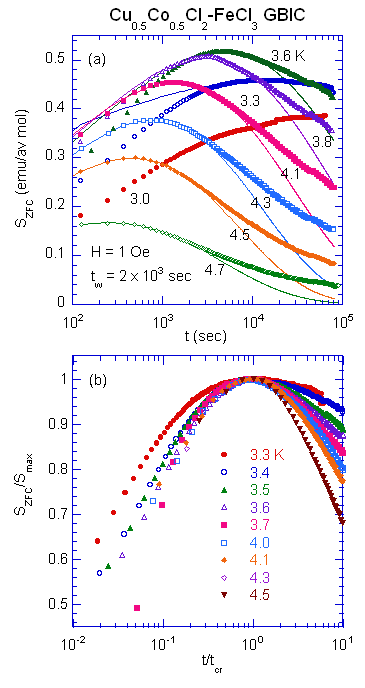}
\caption{\label{fig02}(Color online) (a) $t$ dependence of $S_{ZFC}(t)$ at various $T$ for Cu$_{0.5}$Co$_{0.5}$Cl$_{2}$-FeCl$_{3}$ GBIC. $3.0\le T\le 4.7$ K $t_{w}=2.0\times 10^{3}$ sec. $H$ = 1 Oe. The solid lines denote the least-squares fits to a SER given by Eq.(\ref{EQN11}) for $S_{ZFC}$ vs $t$ in the vicinity of the peak time $t_{cr}$. The fitting parameters $S_{max}^{0}$, $b$, and $\tau_{SER}$ in Eq.(\ref{EQN11}) thus determined are shown in Figs.~\ref{fig03}(a), (b), and (c) as a function of $T$. $\tau_{SER}$ is the relaxation time for the SER. $b$ is the SER exponent. (b) Scaling plot of the ratio $S_{ZFC}(t)/S_{max}$ as a function of $t/t_{cr}$ for $3.3\le T\le 4.5$ K. The values of $t_{cr}$ and $S_{max}$ are shown in Figs.~\ref{fig03}(a) and (c), respectively. $H$ = 1 Oe. $t_{w}=2.0\times 10^{3}$ sec.}
\end{figure}

We have measured the $t$ dependence of $M_{ZFC}(t)$ at various $T$ after the ZFC aging protocol where $H$ = 1 Oe and $t_{w}=2.0\times 10^{3}$ sec.\cite{ref07,ref08,ref09} Figure \ref{fig02}(a) shows the $t$ dependence of $S_{ZFC}(t)$ at various $T$, where $H$ = 1 Oe and $t_{w}=2.0\times 10^{3}$ sec. The relaxation rate $S_{ZFC}(t)$ shows a peak (the peak height $S_{max}$) at the peak time $t=t_{cr}$, which shifts to the short-$t$ side with increasing $T$. The peak time $t_{cr}$ drastically decreases with increasing $T$ near $T=T_{SG}$. The existence of the peak at $t=t_{cr}$ in $S_{ZFC}(t)$ vs $t$ indicates that the SER plays a significant role around $t=t_{cr}$. We find that $S_{ZFC}(t)$ is well described by the SER form given by Eq.(\ref{EQN11}). The least- squares fit of these data of $S_{ZFC}(t)$ vs $t$ to Eq.(\ref{EQN11}) yields the parameters $b$, $\tau_{SER}$, and $S_{max}^{SER}$. Figure \ref{fig02}(b) shows the plot of $S_{ZFC}(t)/S_{max}$ as a function of $t/t_{cr}$ at various $T$, where $S_{max}$ and $t_{cr}$ are different for different $T$. 

\begin{figure}
\includegraphics[width=7.0cm]{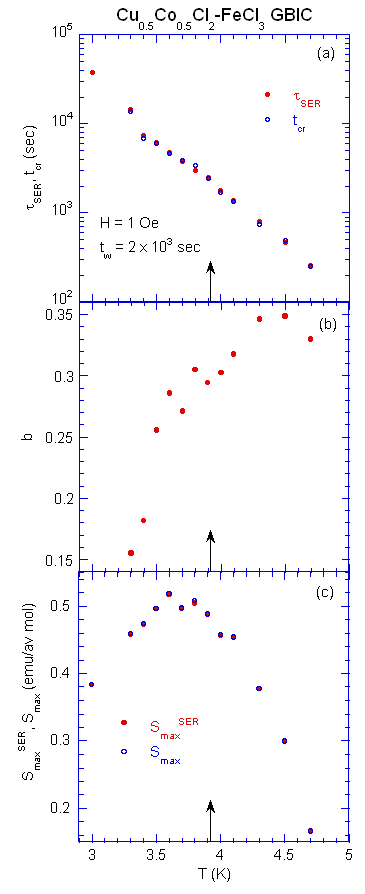}
\caption{\label{fig03}(Color online) (a) $t_{cr}$ vs $T$ and $\tau_{SER}$ vs $T$. (b) $b$ vs $T$. (c) $S_{max}^{SER}$ vs $T$ and $S_{max}$ vs $T$. $H$ = 1 Oe. $t_{w}= 2.0\times 10^{3}$ sec. $t_{cr}$ is a characteristic time at which $S_{ZFC}(t)$ exhibits a peak at the fixed $T$. $S_{max}$ is the peak height of $S_{ZFC}(t)$ at $t=t_{cr}$. The arrows indicates the location of $T_{SG}$ (= 3.92 K).}
\end{figure}
 
Figures \ref{fig03}(a), (b), and (c) show the $T$ dependence of $\tau_{SER}$, $b$, and $S_{max}^{SER}$, determined from the least-squares fits, where $T_{SG} = 3.92$ K, $t_{w}=2.0\times 10^{3}$ sec and $H$ = 1 Oe. The values of $t_{cr}$ and $S_{max}$ are also plotted as a function of $T$. Note that the peak time $t_{cr}$ and the height $S_{max}$ are derived directly from the $t$ dependence of $S_{ZFC}(t)$. We find that $\tau_{SER}$ drastically decreases with increasing $T$ in the vicinity of $T_{SG}$. The value of $\tau_{SER}$ at each $T$ is almost the same as that of $t_{cr}$ at the same $T$. It should be noted that $t_{cr}$ (or $\tau_{SER}$) is equal to $t_{w}$ at $T=T_{SG}$. In Fig.~\ref{fig03}(b) we show the $T$ dependence of $b$. The exponent $b$ is equal to 0.15 at $T=3.3$ K and increases with increasing $T$. The exponent $b$ is nearly equal to 0.3 at $T=T_{SG}$. This value of $b$ is in good agreement with the value at $T_{SG}$ which is predicted by Ogielski\cite{ref12} from the Monte Carlo simulation. Similar behavior has been observed by Bontemps and Orbach\cite{ref14} and Bontemps\cite{ref15} in the curve $b$ vs $T$ for the insulating spin glass Eu$_{0.4}$Sr$_{0.6}$S ($T_{SG}$ = 1.5 K): the exponent $b$ is nearly equal to 0.17 at 1.3 K. It increases with increasing $T$. It is equal to 0.3 at $T=T_{SG}$ and increases to 0.4 just above $T_{SG}$.

In Fig.~\ref{fig03}(c) we show the values of $S_{max}$ and $S_{max}^{SER}$ as a function of $T$. The value of $S_{max}$ is almost equal to that of $S_{max}^{SER}$ at the same $T$. We find that $S_{max}$ shows a broad peak at $T$ = 3.6 K just below $T_{SG}$. The deviation of the data of $S_{ZFC}(t)$ vs $t$ from the SER occurs at both $t\gg t_{cr}$ and $t\ll t_{cr}$. For convenience, in Fig.~\ref{fig02}(b) we define characteristic times $t_{u}$ ($>t_{cr}$) and $t_{l}$ ($<t_{cr}$), where $S_{ZFC}(t)$ reaches a $0.8 S_{max}$; $\ln(t_{u}/t_{cr})=0.6012/b$ and $\ln(t_{l}/t_{cr})=-0.7515/b$. The times $t_{u}$ and $t_{l}$ approach the peak time $t_{cr}$ as $b$ is increased. It follows that the width of the curve $S_{ZFC}(t)/S_{max}$ vs $t/t_{cr}$, defined by $\ln (t_{u}/t_{l})=1.35/b$, becomes narrower as $b$ is increased from $b_{0}=4a=0.14$ to $b = 1$. The width of the curve ($S_{ZFC}(t)/S_{max}$ vs $t$) becomes narrower with increasing $T$. This implies that $b$ increases with increasing $T$ (see Fig.~\ref{fig03}(b) for comparison).

\begin{figure}
\includegraphics[width=7.5cm]{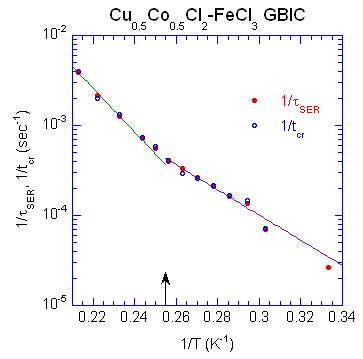}
\caption{\label{fig04}(Color online) Plot of $1/\tau_{SER}$ and $1/t_{cr}$ as a function of $1/T$. $H$ = 1 Oe. $t_{w}= 2.0\times 10^{3}$ sec. The solid lines denote the best fit of the SER form given by Eq.(\ref{EQN11}) to experimental data for $T<T_{SG}$ and $T>T_{SG}$, respectively. The arrow indicates the location of $T_{SG}$.}
\end{figure}

In Fig.~\ref{fig04} we show the plot of $1/\tau_{SER}$ (or $1/t_{cr}$) as a function of $1/T$. There is a drastic change in $1/\tau_{SER}$ vs $1/T$ around $1/T=1/T_{SG}=0.255$ K$^{-1}$. It is expected that the $t$ dependence of $\tau_{SER}$ (or $t_{cr}$) is given by an Arrhenius law,
\begin{equation} 
1/\tau_{SER}=c_{0}^{*}\exp(-c_{1}^{*}T_{SG}/T),
\label{EQN17} 
\end{equation} 
with different $c_{0}^{*}$ and $c_{1}^{*}$ for $T>T_{SG}$ and $T<T_{SG}$. The least-squares fit of our data of of $1/\tau_{SER}$ (or $1/t_{cr}$) yields the parameters $c_{0}^{*}= 19.7\pm 11.5$ sec$^{-1}$ and $c_{1}^{*}=10.82\pm 0.56$ for $T>T_{SG}$, and $c_{0}^{*}=1.01 \pm 0.48$ sec$^{-1}$ and $c_{1}^{*}=7.81\pm 0.46$ for $T<T_{SG}$. The temperature corresponding to the characteristic energy barrier of the relaxation process is $E_{B}=c_{1}^{*}T_{SG}=42.4$ K for $T>T_{SG}$ and 30.6 K for $T<T_{SG}$. The same form of the $1/\tau_{SER}$ vs $1/T$ has been used by Hoogerbeets et al.\cite{ref16} for their analysis of TRM relaxation measurements of canonical SG systems: Ag:Mn (2.6 at. \%), Ag: Mn (4.1 at. \%), Ag:[Mn (2.6 at. \%) + Sb (0.46 at. \%)], and Cu:Mn (4.0 at \%). Our value of $c_{1}^{*}$ is much larger than those derived by Hoogerbeets et al.\cite{ref16} ($c_{1}^{*} = 2.5$). Note that in their work the stretched exponential was taken as representative of the short time ($t<t_{w}$) relaxation.

\subsection{\label{resultB}Aging behavior at various $t_{w}$}

\begin{figure}
\includegraphics[width=7.0cm]{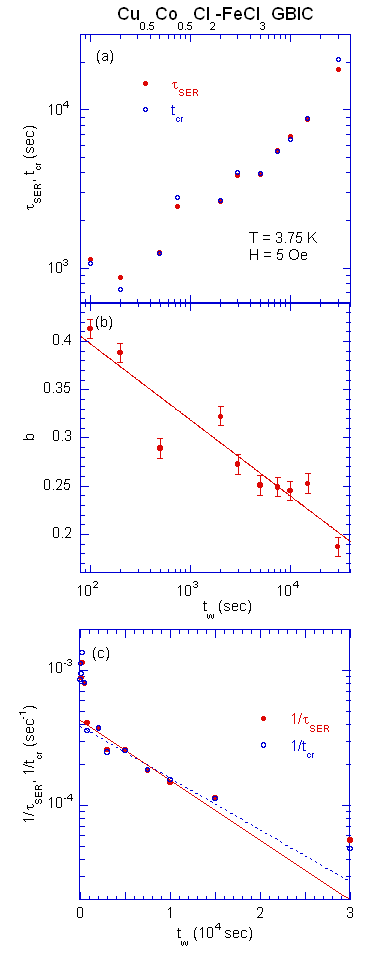}
\caption{\label{fig05}(Color online) (a) $t_{cr}$ vs $t_{w}$ and $\tau_{SER}$ vs $t_{w}$. (b) the SER exponent $b$ vs $t_{w}$. $T$ = 3.75 K. $H$ = 5 Oe. $1.0\times 10^{2}\le t_{w}\le 3.0\times 10^{4}$ sec. The solid line is the best fit of Eq.(\ref{EQN18}) to the data of $b$ vs $t_{w}$. These results are obtained from the measurement of $M_{ZFC}(t)$ as a function of $t$ after the ZFC cooling protocol and isothemal aging at $T$ (= 3.75 K) for a wait time $t_{w}$. (c) $1/\tau_{SER}$ vs $t_{w}$ and $1/t_{cr}$ vs $t_{w}$. $T$ = 3.75 K. $H$ = 5 Oe. The solid line is the best fit of Eq.(\ref{EQN19}) to the data of $1/\tau_{SER}$ vs $t_{w}$.}
\end{figure}

We have measured the $t$ dependence of $M_{ZFC}(t)$ at $T$ = 3.75 K and $H$ = 5 Oe just below $T_{SG}$ after the ZFC aging protocol, where the system was aged at $T$ = 3.75 K for the wait time $t_{w}$. This wait time $t_{w}$ is varied as a parameter: $1.0\times 10^{2}\le t_{w}\le 3.0\times 10^{4}$ sec. A least-squares fit of these data in the vicinity of $t=t_{cr}$ to Eq.(\ref{EQN11}) for $t/t_{cr}<10$ yields the parameters $\tau_{SER}$ and $b$ for each $t_{w}$. Figures \ref{fig05}(a) and (b) show the plot of $\tau_{SER}$, $t_{cr}$, and $b$ as a function of $t_{w}$. We find that the exponent $b$ decreases from 0.4 to 0.2 with increasing $t_{w}$ from $1.0\times 10^{2}$ sec to $3\times 10^{4}$ sec. The data of $b$ vs $t_{w}$ for $1.0\times 10^{2}\le t_{w}\le 3.0\times 10^{4}$ sec is described by 
\begin{equation} 
b=b_{0}^{*}-b_{1}^{*}\ln (t_{w}/t_{w}^{0}),
\label{EQN18} 
\end{equation} 
where $t_{w}^{0}$ is chosen as $1.0\times 10^{2}$ sec, $b_{0}^{*}=0.40\pm 0.02$, and $b_{1}^{*}=0.034\pm 0.004$. Using Eq.(\ref{EQN18}), the value of $b$ can be estimated as $b$ = 0.14 at $t_{w}=2.0\times 10^{5}$ sec, which is nealy equal to $4a$. 

In Fig.~\ref{fig05}(c) we show the plot of $1/\tau_{SER}$ as a function of $t_{w}$. It is suggested by Chamberlin.\cite{ref17} that $\tau_{SER}$ is described by a form
\begin{equation} 
1/\tau_{SER}=\omega^{*}\exp(-t_{w}/t_{\omega}^{*}),
\label{EQN19} 
\end{equation} 
where $\omega^{*}$ and $t_{w}^{*}$ are constant to be determined. A least-squares fit of the data of $1/\tau_{SER}$ vs $t_{w}$ in the limited $t_{w}$-region $750\le t_{w}\le 1.5\times 10^{4}$ sec to Eq.(\ref{EQN19}) yields the parameters $\omega_{0}^{*} = (3.86 \pm 0.25)\times 10^{-4}$ sec$^{-1}$ and $t_{w}^{*}=(1.13 \pm 0.18)\times 10^{4}$ sec. The values of $1/\tau_{SER}$ for $t_{w}<750$ sec considerably deviates from the exponential $t_{w}$ dependence, partly because of the initial stage of the aging process depending on the initial condition. 

\subsection{\label{resultC}Aging behavior under the $T$-shift}

\begin{figure}
\includegraphics[width=7.5cm]{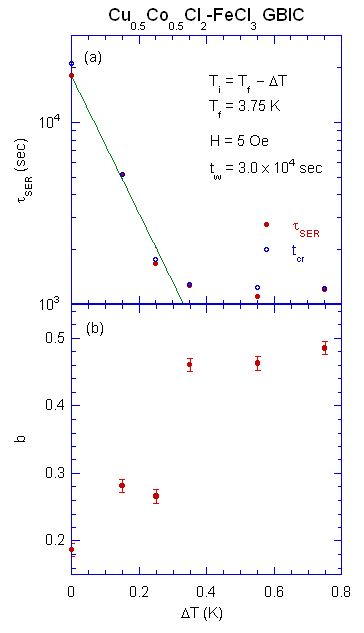}
\caption{\label{fig06}(Color online) (a) $t_{cr}$ vs $\Delta T$ and $\tau_{SER}$ vs $\Delta T$. The solid line denotes the best fit of Eq.(\ref{EQN20}) to the data of $\tau_{SER}$ vs $\Delta T$ for $0\le \Delta T\le 0.25$ K. (b) $b$ vs $\Delta T$. $H$ = 5 Oe. $t_{w}=3.0\times 10^{4}$ sec. The results are obtained from the measurement of $\chi_{ZFC}(t)$ as a function of $t$, after the ZFC cooling protocol, isothermal aging at $T=T_{i}=T_{f}-\Delta T$. Immediately after the temperature is shifted to $T=T_{f}$ (so-called $\Delta T$ shift) and the magnetic field is turned on, $M_{ZFC}$ is measured as a function of $t$.}
\end{figure}

\begin{figure*}
\includegraphics[width=14.0cm]{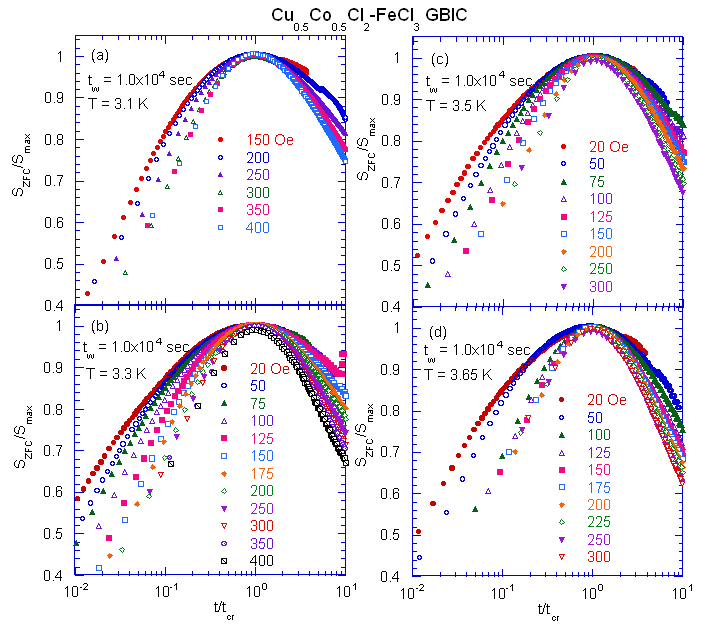}
\caption{\label{fig07}(Color online) Scaling plot of the ratio $S_{ZFC}(t)/S_{max}$ as a function of $t/t_{cr}$. $t_{w}=1.0\times 10^{4}$ sec. $H$ is changed as a parameter. (a) $T$= 3.1 K, (b), 3.3 K, (c) 3.5 K, and (d) 3.65 K.}
\end{figure*}

We have measured the $t$ dependence of $M_{ZFC}(t)$ under the $T$-shift from the initial temperature $T_{i}=T-\Delta T$ to the final temperature $T_{f}=T$, where $T_{i}$ = 3, 3.2, 3.4, 3.5, 3.6, and 3.9 K. In the $T$-shift experiment, the system is cooled in zero field from 50 K to a temperature $T_{i}=T-\Delta T$ below $T_{SG}$. After a wait time $t_{w}$ ($= 3.0\times 10^{4}$ sec) at this temperature, immediately prior to the field application, the temperature is raised to $T_{f}$ = 3.75 K. The $t$ dependence of $M_{ZFC}(t)$ was measured at $H$ = 5 Oe. We have already reported the $t$ dependence of $S_{ZFC}(t)$ under the $T$-shift in our previous paper.\cite{ref08} We find that $S_{ZFC}(t)$ shows a peak at $t_{cr}$. The peak shifts to the long-$t$ side as $\Delta T$ is decreased. A least-squares fit of the data $S_{ZFC}(t)$ vs $t$ in the vicinity of $t=t_{cr}$ to Eq.(\ref{EQN11}) yields the parameters $\tau_{SER}$ and $b$. In Figs.~\ref{fig06}(a) and (b), we show the parameters $\tau_{SER}$ (also $t_{cr}$) and $b$ as a function of $\Delta T$. The relaxation time $\tau_{SER}$  decrease with increasing $\Delta T$, while $b$ increases with increasing $\Delta T$. 

In our previous paper,\cite{ref08} we have shown an expression for $t_{cr}$ under the $\Delta T$-shift aging process. This expression is derived from the Monte Carlo simulation based on the droplet model (Takayama and Fukushima\cite{ref18}) under the condition that the domain size is comparable to the overlap length $L_{\Delta T}$. In the limit of $\Delta T\rightarrow 0$, $\ln \tau_{SER}$ is linearly dependent on $\Delta T$,
\begin{equation} 
\tau_{SER}=\tau_{T}^{*}\exp(-\alpha_{T}^{*}\Delta T),
\label{EQN20} 
\end{equation} 
where a slope $\alpha_{T}^{*}$ and a relaxation time $\tau_{T}^{*}$ are to be determined. The slope $\alpha_{T}^{*}$ increases with increasing $t_{w}$. The curve ($\ln\tau_{SER}$ vs $\Delta T$) for $0\le\Delta T\le 0.25$ K is linearly dependent on $\Delta T$ (see Fig.~\ref{fig06}(a)). The fitting parameters are given by slope $\alpha_{T}^{*}= 8.7\pm 0.5$ K$^{-1}$ and $\tau_{T}^{*}=(1.81\pm 0.05)\times 10^{4}$ sec.

\subsection{\label{resultD}Aging behavior under the $H$-shift}

\begin{figure}
\includegraphics[width=7.5cm]{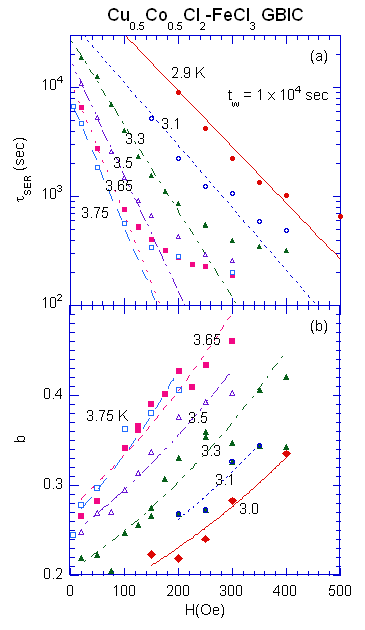}
\caption{\label{fig08}(Color online) (a) $\tau$ vs $H$ at various $T$. $t_{w} = 1.0\times 10^{4}$ sec. The lines denote the best fits of Eq.(\ref{EQN21}) to the experimental data in the low $H$ limit. (b) $b$ vs $H$. The lines are guides to the eyes. $t_{w}= 1.0\times 10^{4}$ sec. The results are obtained from the measurement of $M_{ZFC}(t)$ as a function of $t$ in the presence of $H$, after the ZFC cooling protocol and isothermal aging at $T$ for a wait time $t_{w}$ at $H$ = 0.}
\end{figure}

\begin{figure}
\includegraphics[width=7.5cm]{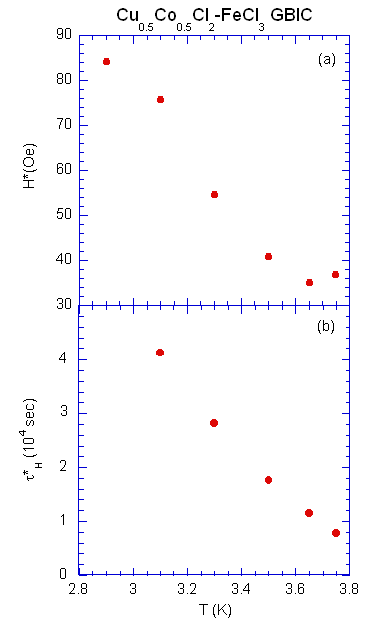}
\caption{\label{fig09}(Color online) $T$ dependence of fitting parameters $H^{*}$ and $\tau_{H}^{*}$ determined from the least-squares fit of the data of Fig.~\ref{fig08}(a) to Eq.(\ref{EQN21}). (a) $H^{*}$ vs $T$. (b) $\tau_{H}^{*}$ vs $T$.}
\end{figure}

We have measured the $t$ dependence of $M_{ZFC}(t)$ after the ZFC aging protocol which consists of (i) annealing at 50 K for $1.2\times 10^{3}$ sec at $H$ = 0, (ii) cooling from 50 K to $T$ ($<T_{SG}$), (iii) isothemal aging at $T$ for a wait time $t_{w}$ ($= 1.0\times 10^{4}$ sec), and (iv) switching a field from 0 to $H$. The time $t$ = 0 is a time when $H$ is turned on. Our results on the dependence of $S_{ZFC}(t)$ at the fixed $T$ and $H$ of the $(H,T)$ plane, have been reported in our previous paper.\cite{ref09} The relaxation rate $S_{ZFC}(t)$ exhibits a peak [the peak height $S_{max}(T,H)$] at a peak time $t_{cr}(T,H)$, which drastically shifts to the short-$t$ side with increasing $H$. Figure \ref{fig07} shows the scaling plot of $S_{ZFC}(t)/S_{max}$ as a function of $t/t_{cr}$ at $T$ = 3.1, 3.3, 3.5, and 3.65 K. The magnetic field is changed as a parameter. The least-squares fit of the data of $S_{ZFC}(t)$ vs $t$ in the vicinity of $t_{cr}(T,H)$ to the SER given by Eq.(\ref{EQN11}), yields the parameters $\tau_{SER}$ and $b$. In Figs.~\ref{fig08}(a) we show the $H$ dependence of $\tau_{SER}$ at various $T$ below $T_{SG}$. Note that the values of $\tau_{SER}$ and $t_{cr}$ are almost the same at the same $T$ and $H$. We find that $\tau_{SER}$ drastically decreases with increasing $H$ at each $T$ below $T_{SG}$. In Fig.~\ref{fig08}(b) we show the $H$ dependence of $b$ at various $T$. The exponent $b$ increases with increasing $H$ at each $T$ and reaches a value between 0.4 and 0.5. The field $H$ at which $b$ is equal to 0.3, increases with decreasing $T$ from 3.75 to 2.9 K. Note that $b$ decreases with increasing the cooling field $H_{c}$ just below $T_{SG}$ for the TRM decay experiment on Ag:Mn(2.6 at \%) + Sb(0.46 at \%) (Hoogerbeets et al.\cite{ref16}) and Cu: Mn (6.0 at \%) (Chu et al.\cite{ref19}). This result is different from our result from the ZFC magnetization relaxation measurement that $b$ increases with increasing the applied field $H$. 

In our previous paper,\cite{ref08} we have shown an expression for $t_{cr}$ under the $H$-shift aging process. This expression is derived from the Monte Carlo simulation based on the droplet model (Takayama and Fukushima\cite{ref20}) under the condition that the domain size is comparable to the crossover length $L_{H}$. In the limit of $H\rightarrow 0$. $\ln t_{cr}$ is linearly dependent on $H$: $\ln t_{cr} \approx -\alpha_{H}^{*}H$, where $\alpha_{H}^{*}$ is constant. We assume that $\tau_{SER}$ is described by the form
\begin{equation} 
\tau_{SER}=\tau_{H}^{*}\exp(-H/H^{*}),
\label{EQN21} 
\end{equation} 
in the low-$H$ limit, where $H^{*}=1/\alpha_{H}^{*}$. In Fig.~\ref{fig08}(a) we show the $H$ dependence of $\tau_{SER}$ at various $T$ below $T_{SG}$ where $t_{w}=1.0\times 10^{4}$ sec. We find that $\ln \tau_{SER}$ is proportional to $H$ at low $H$. The slope $\alpha_{H}^{*}$ gradually decreases as $T$ is lowered for $T<T_{SG}$. The least-squares fit of the data of $\tau_{SER}$ vs $H$ at low $H$ to Eq.(\ref{EQN21}) yields parameters $\tau_{H}^{*}$ and $H^{*}$ at each $T$. In Fig.~\ref{fig09} we show the $T$ dependence of $H^{*}$ and $\tau_{H}^{*}$. The characteristic field $H^{*}$ decreases with increasing $T$ below $T_{SG}$. The characteristic relaxation time $\tau_{H}^{*}$ decreases with increasing $T$ and reduces to zero around $T=T_{SG}$. 

\section{\label{dis}DISCUSSION}

\begin{figure}
\includegraphics[width=7.5cm]{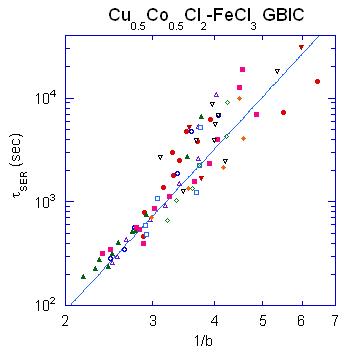}
\caption{\label{fig10}(Color online) Relation of $\tau_{SER}$ vs $1/b$. All the data of $\tau_{SER}$ vs $b$ obtained in the present work are plotted. The data fall well on a single curve given in the text, irrespective of the values of $t_{w}$, $H$, $T$, and $\Delta T$. The data with different notations are obtained under different conditions ($t_{w}$, $T$, $H$, and $\Delta T$).}
\end{figure}

\begin{figure}
\includegraphics[width=7.5cm]{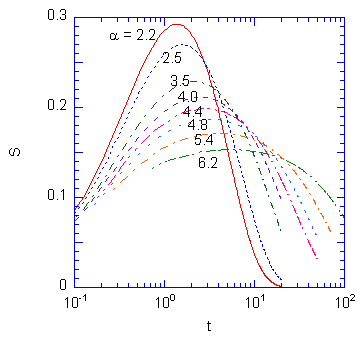}
\caption{\label{fig11}(Color online) Relaxation rate $S(t)$ vs $t$ at various values of $\alpha$: The parameter $1/\alpha$ is proportional to $T$. The $t$ dependence of $S(t)$ is derived from numerical calculation of the differential equation given by Eq.(\ref{EQN24}). The relaxation rate $S(t)$ is well described by the SER form defined by Eq.(\ref{EQN11}) with $b$ in the vicinity of the peak time.}
\end{figure}

\begin{figure*}
\includegraphics[width=14.0cm]{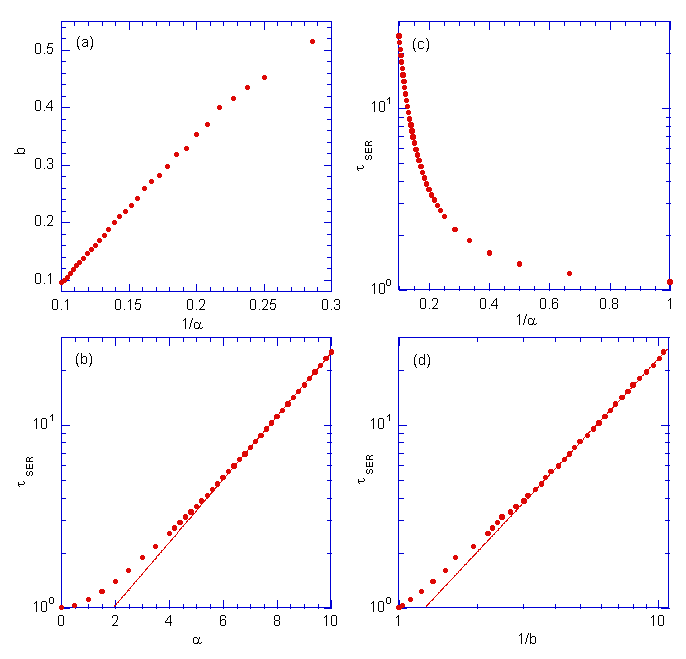}
\caption{\label{fig12}(Color online) Fitting parameters of the SER. (a) $b$ vs $1/\alpha$, where $1/\alpha$ is proportional to $T$. The exponent $b$ is equal to 0.3 at $1/\alpha$ = 0.175. (b) $\tau_{SER}$ vs $\alpha$, which denotes the Arrhenius law. (c) $\tau_{SER}$ vs $1/\alpha$. The relaxation time $\tau_{SER}$ increases with decreasing $1/\alpha$. (d) The relationship between $\tau_{SER}$ and $1/b$. The solid lines in (b) and (d) denote the best fitting curves (see the text).}
\end{figure*}

We have shown that $S_{ZFC}(t)$ exhibits a peak at a peak time $t_{cr}$. The relaxation rate $S_{ZFC}(t)$ is well described by a SER in the vicinity of $t_{w}$. The relaxation time $\tau_{SER}$ obeys the Arrhenius law. The exponent $b$ for the SER increases with increasing $T$ and is nearly equal to 0.30 at $T=T_{SG}$. This value of $b$ at $T_{SG}$ is in good agreement with the prediction from the numerical simulation on the $\pm J$ Ising spin glasses (Ogielski\cite{ref12}). The exponent $b$ increases with increasing $T$ above $T_{SG}$ and tends to approach 1 (exponential type relaxation, Debye-type) well above $T_{SG}$. Similar conclusion has been derived by Keren et al.\cite{ref21} from the $T$ dependence of the spin-spin dynamical autocorrelation function of the Ising SG, Fe$_{0.05}$TiS$_{2}$: $b$ is equal to 1/3 at $T=T_{SG}$ and $b$ increases with increasing $T$ above $T_{SG}$. The value of $b\approx 1/3$ at $T=T_{SG}$ may imply that the available configuration space tends to the structure of a percolation fractal, so the SG transition is physically a percolation in phase space (Almeida et al.\cite{ref22}) 

Theoretically it is predicted that the exponent $b$ is related to the critical exponents by\cite{ref23,ref24} 
\begin{equation}
b=\frac{\beta +\gamma }{\beta +\gamma +z\nu } ,
\label{EQN22}
\end{equation}
where $\gamma$ is the exponent of the susceptibility. Using our critical exponents ($\beta=0.36\pm 0.03$. $\gamma =3.5\pm 0.4$ , and $z\nu=10.3\pm 0.7$), the critical exponent $b$ for Cu$_{0.5}$Co$_{0.5}$Cl$_{2}$-FeCl$_{3}$ GBIC can be estimated as $b$ = 0.27, which is close to 0.3 at $T=T_{SG}$. Note that the criterion ($b>4a$) is satisfied for the existence of the peak of $S_{ZFC}(t)$ at $t=t_{cr}$, since $a=0.035$. 

Figure \ref{fig10} shows the plot of $\tau_{SER}$ vs $1/b$ for all the data obtained in the present work, including Figs.~\ref{fig03}(a) and (b), Figs.~\ref{fig05}(a) and (b), Figs.~\ref{fig06}(a) and (b), and Figs.~\ref{fig08}(a) and (b). The data of $\tau_{SER}$ vs $1/b$ for $2<1/b<3.3$ (corresponding to $0.3<b<0.5$) fall, to within experimental accuracy, on a single line which is the best fit of the form,
\begin{equation} 
\ln\tau_{SER}=p_{0}^{*}+p_{1}^{*}\ln (1/b),
\label{EQN23} 
\end{equation} 
with $p_{0}^{*}= 0.9\pm 0.3$, $p_{1}^{*}=5.2\pm 0.5$. On the other hand, the data of $\tau_{SER}$ vs $1/b$ for $3.3<1/b <7$ ($0.1<b<0.3$) are rather broadly distributed near the single line. This result suggests that the exponent $b$ is closely related to the SER time $\tau_{SER}$, irrespective of the values of $t_{w}$, $T$, $H$, and $\Delta T$. Note that Hoogerbeets et al.\cite{ref16,ref25} have reported a relationship between $\ln \tau_{SER}$ and $1/b$ below $T_{SG}$ for canonical SG systems: Ag:Mn (2.6 at. \%), Ag: Mn (4.1 at. \%), Ag:[Mn (2.6 at. \%) + Sb (0.46 at. \%)], and Cu:Mn (4.0 at \%): They have shown that $\ln \tau_{SER}$ decreases with increasing $1/b$. Their results are very different from our result. The correlation between $\tau_{SER}$ and $b$ has been studied by Goltzer et al.\cite{ref26} using Monte Carlo simulations on the Ising SG's: $\tau_{SER}$ increases and $b$ decreases as $T$ is approached $T_{SG}$ from the high-temperature side. This result is similar to our result. 

A successful theory of the glassy relaxation should provide a justification for two common features: (i) the SER form of the relaxation rate and (ii) the correlation between $\tau_{SER}$ and $1/b$. These features are seen in many systems other than SG's, and is considered to be a signature of the glassy relaxation. Here we consider a simple model proposed by Trachenko and Dove,\cite{ref10} and Trachenko\cite{ref11} for the glassy transitions. The dynamics of the local relaxation events is governed by a differential equation with a solution that fits well to the stretched-exponential relaxation. The rate equation is given by
\begin{equation} 
\frac{dx(t)}{dt}=\exp[-\alpha x(t)]-x(t)\exp(-\alpha ),
\label{EQN24} 
\end{equation} 
with an initial condition $x(0)=0$, where $x(t)$ may correspond to the ratio $M_{ZFC}(t)/M_{ZFC}(t=\infty)$. The parameter $\alpha$ is proportional to $E_{B}/k_{B}T$, where $E_{B}$ is the activation barrier, $k_{B}$ is the Boltzmann constant. The time $t$ is redefined as $t/t_{0}$, where $t_{0}$ is a characteristic time. The second term of Eq.(\ref{EQN24}) describes saturation, such that $dx(t)/dt=0$ as $t\rightarrow \infty$, or $x(t) \rightarrow 1$. The relaxation rate $S(t)$ is defined by
\begin{equation} 
S(t)=\frac{dx(t)}{d\ln t}=t\{\exp [-\alpha x(t)]-x(t)\exp(-\alpha )\}.
\label{EQN25} 
\end{equation} 
Figure \ref{fig11} shows the result of numerical calculation of $S(t)$, where $\alpha$ is changed as a parameter ($\alpha = 0 - 10$). The parameter ($1/\alpha$) is proportional to $T$. The relaxation rate $S(t)$ has a peak around $t=1$. This peak shifts to longer-$t$ side with decreasing ($1/\alpha$) (or with decreasing $T$). We find that $S(t)$ is well described by the SER form given by Eq.(\ref{EQN11}) in the vicinity of the peak time. The least-squares fit of the data of $S(t)$ vs $t$ to the SER form given by Eq.(\ref{EQN11}) yields the parameters $b$ and $\tau_{SER}$. Figures \ref{fig12}(a) shows the plot of $b$ vs $1/\alpha$. The exponent $b$ increases from $b$ = 0.1 to 0.4 with increasing ($1/\alpha$). The exponent $b$ is equal to 0.3 around $1/\alpha=1/\alpha_{SG} = 0.175$ corresponding to $T_{SG}$ ($\alpha_{SG}=5.71$). Figure \ref{fig12}(b) shows the relaxation time $\tau_{SER}$ as a function of $\alpha$ ($\propto 1/T$), showing the Arrhenius form. The relaxation time $\tau_{SER}$ for $\alpha_{SG}<\alpha<10$ is well described by the form, $\ln \tau_{SER}=d_{0}^{*}+d_{1}^{*}\alpha$ with $d_{0}^{*}=-0.863\pm 0.006$ and $d_{1}^{*}=0.397\pm 0.002$. The parameter $c_{1}^{*}$ for the Arrhenius law Eq.(\ref{EQN17}) is related to by $c_{1}^{*}= d_{1}^{*}\alpha_{SG}=2.26$. This value of $c_{2}^{*}$ is close to that reported by Hoogerbeets et al.\cite{ref16} ($c_{2}^{*}=2.5$). Figure \ref{fig12}(c) shows the relaxation time $\tau_{SER}$ as a function of $1/\alpha$ ($\propto T$). The relaxation time $\tau_{SER}$ drastically increases with decreasing $1/\alpha$ below $1/\alpha=1/\alpha_{SG}$. Figure \ref{fig12}(d) shows the relationship between $\tau_{SER}$ and $1/b$. The relaxation time $\tau_{SER}$ is uniquely determined from the value of $1/b$ by the relation
\begin{equation} 
\ln\tau_{SER}=q_{0}^{*}+q_{1}^{*}\ln(1/b),
\label{EQN26} 
\end{equation} 
for $3.3<1/b<10$, where $q_{0}^{*}= -0.35\pm 0.02$ and $q_{1}^{*}=1.515\pm 0.01$. Our experimental value of $p_{1}^{*}$ ($= 5.2\pm 0.5$) is much larger than the theoretical value of $q_{1}^{*}$, suggesting the incompleteness of the model. In spite of such difference it can be concluded from this model that $\tau_{SER}$ increases with increasing $1/b$. This is the same conclusion derived by Ngai and Tsang.\cite{ref27} In summary, the features of parameters of the SER, can be explained in terms of the simple model of the glassy dynamics.\cite{ref10} 

\section{CONCLUSION}
The aging behavior of Cu$_{0.5}$Co$_{0.5}$Cl$_{2}$-FeCl$_{3}$ graphite bi-intercalation compound has been studied from the time dependence of the relaxation rate $S_{ZFC}(t)$. The relaxation rate $S_{ZFC}(t)$ is described by a SER form in the vicinity of $t=t_{cr}$. There is a correlation between the exponent $\tau$ and $1/b$ of the SER, irrespective of the values of $t_{w}$, $T$, $H$, and $\Delta T$. The exponent $b$ is equal to 0.3 at $T=T_{SG}$. The exponent $b$ increases with increasing $T$. The relaxation time $\tau_{SER}$ obeys the Arrhenius law. These features, which are a signature of the glass relaxation of many systems, can be well explained in terms of the simple relaxation model. 

\begin{acknowledgments}
We would like to thank H. Suematsu for providing us with single crystal kish graphite, and T. Shima and B. Olson for their assistance in sample preparation and x-ray characterization.
\end{acknowledgments}

\end{document}